\documentstyle[proceedings,psfig]{crckapb}

\newcommand{\sect}[1]{\S$\,$\ref{#1}}
\newcommand{\thOr}{$\theta^1\,$Ori~C}
\newcommand{\zero}{_0}
\newcounter{ioncounter}
\newcommand{\ion}[2]{\setcounter{ioncounter}{#2}#1$\,${\sc
  \roman{ioncounter}}} 
\newcommand{\kms}{\,\mbox{km s$^{-1}$}}
\newcommand{\Othree}{[\ion{O}{3}]~5007\AA}
\renewcommand{\P}[1]{%
\ifnum#1=1\hbox{OW~168--326E}\fi
\ifnum#1=2\hbox{OW~167--317}\fi
\ifnum#1=3\hbox{OW~163--317}\fi
\ifnum#1=5\hbox{OW~158--323}\fi
\ifnum#1=0\hbox{OW~171--334}\fi}

\begin{opening}
  \title{TWO-WIND INTERACTION MODELS OF THE PROPLYDS IN THE ORION NEBULA}
  \author{W. J. Henney}
  \author{$\,$S. J. Arthur}
  \institute{Instituto de Astronom\'\i a, UNAM, Unidad Morelia,\\
    J. J. Tablada 1006, 58090 Morelia, Michoac\'an, M\'exico}
\end{opening}
\runningtitle{TWO-WIND INTERACTION MODELS OF PROPLYDS}

\begin{document}

\begin{abstract}
  Many low-mass stars in the Orion nebula are associated with very
  compact ($\simeq\!1\,$arcsec) emission knots, known variously as
  proplyds, PIGs or LV knots. Some of these knots are teardrop-shaped,
  with ``tails'' pointing away from the massive star \thOr, which is
  the principal exciting star of the nebula. We discuss models of such
  knots, which invoke the interaction of the fast stellar wind from
  \thOr\ with a transonic photoevaporated flow from the surface of an
  accretion disk around a young low-mass star. We review previous
  analytic work and compare the results of the model with the observed
  brightnesses, morphologies and emission line profiles of the knots,
  as well as presenting new results from numerical hydrodynamical
  simulations.
\end{abstract}

\section{Introduction}
\label{sec:intro}
The proplyds are bright compact emission line knots, with sizes of
order 0.5--2.0 arcseconds, that are found in the inner region of the
Orion nebula \cite{Laq79,Gar87,Chu87,Fel93,ODe93,ODe94,p:McC97} and
nearly all of which contain an embedded low-mass star
\cite{Mea88c,McC94}. Many of the proplyds show a head/tail morphology,
in which the tail points away from the star \thOr, the most massive
star of the Trapezium cluster. Emission line spectroscopy of the
proplyds in the \Othree\ line \cite{Mas93,Mas95,Hen97} show a bright
central core with full width half maximum (FWHM) of $\simeq 50 \kms$,
together with faint wings extending out to $\ge 100 \kms$ from the
line center.

The obvious explanation for these objects is that the radiation and
stellar wind from \thOr\ is interacting with the circumstellar
material around the low-mass stars. In our models we assume that this
circumstellar material is in the form of an optically thick,
geometrically thin accretion disk. The effect of the radiation from
\thOr\ will be to ionize the material in the disk, which will produce
a photoevaporated flow away from the disk and {\em towards} the
ionizing source. Hence, to produce the tails pointing away from \thOr,
this flow must somehow be confined and redirected. Two possible
candidates for this confinement mechanism are the exciting star's
radiation pressure and the ram pressure of its stellar wind. However,
only the second of these is feasible, as is shown in
\sect{sec:constraints}. In \sect{sec:anal} a simple analytic model of
the resultant two-wind interaction is briefly described and the
successes and failures of the model in explaining observed properties
of the proplyds are outlined in \sect{sec:comp}. In \sect{sec:num},
preliminary results of numerical hydrodynamic simulations are
presented, which remove some of the arbitrary assumptions of the
model. Complications such as the possible existence of a neutral
photodissociated flow are critically discussed in \sect{sec:discuss}.

\section{Confinement Mechanisms --- Radiation vs.\ Ram Pressure} 
\label{sec:constraints}
The gas pressure at the base of the ionized flow can be calculated
simply by equating the numerical flux, $F_0$, of Lyman continuum
(Ly-c) photons arriving at the ionization front (IF) with the
numerical flux of newly-ionized ions entering the photoevaporated
wind:
\begin{equation}
  \label{base}
  F\zero = n\zero\, u\zero \ , 
\end{equation}
where $n\zero$ is the ion number density and $u\zero$ the ion velocity
at the base of the wind. This leads to the following expression for
the gas pressure
\begin{equation}
  \label{pgas}
  P_{\rm gas} \equiv \mu m_{\rm H} n\zero\, c\zero^2 = \mu m_{\rm H}
  c\zero F\zero\, / {\cal M}\zero \ , 
\end{equation}
where $\mu$ is the mean atomic mass ($\simeq 1.3$), $m_{\rm H}$ is the
mass of hydrogen, $c\zero$ is the sound speed in the ionized gas
($\simeq 12\kms$) and ${\cal M}\zero$ is the Mach number at the base
of the flow, which will be of order 1--2 \cite{Dys68,Kah69,Ber89}.

The {\em unattenuated} ionizing flux is given by $F_\star =
\dot{S_*}/{4 \pi d^2}$, where $\dot{S_*}$ is the stellar ionizing
photon rate, for which estimates vary between $ 7 \times 10^{48}\, {\rm
  s}^{-1}$ \cite{Pan73} and $3\times 10^{49}\, {\rm s}^{-1}$
\cite{Ber96}, and $d$ is the distance of the proplyd from the exciting
star. However, at the distances of the proplyds from \thOr, most of
this flux is used up in maintaining the ionization state of the
photoevaporated flow against recombination. With the assumption that
$F_0 \ll F_\star$, one can write
\begin{equation}
  \label{ionbalance}
  F_\star e^{-\tau_0} = f(\tau_0)\, n\zero^2 \alpha_{\rm B} r_{\rm d} \ , 
\end{equation}
where $\tau_0$ is the dust absorption optical depth of the flow,
$\alpha_{\rm B}$ is the Case B recombination coefficient ($2.6 \times
10^{-13}$ cm$^3$ s$^{-1}$), $r_{\rm d}$ is the disk radius and
$f(\tau_0) \simeq (3 + \tau_0)^{-1}$ depends slightly on the assumed
geometry \cite{Hen96}. Assuming $\tau_0 = 0$, one finds
(Eqs.~\ref{base} and~\ref{ionbalance}) that the {\em percentage} of
ionizing photons reaching the ionization front is
\begin{equation}
  \label{beta}
  \beta_\% = \frac{100\, F\zero}{F\star} = 1.5\, {\cal M}\zero\, d_{17}\,
  r_{15}^{-1/2}\, \dot{S}_{49}^{-1/2} \ , 
\end{equation}
where $d_{17}$ and $r_{15}$ are the proplyd distance and disk radius
measured in units of $10^{17}$~cm ($\simeq 0.3$~pc) and $10^{15}$~cm
($\simeq 66$~au) respectively and $\dot{S}_{49}$ is the stellar
ionizing photon rate in units of $10^{49}\,{\rm s}^{-1}$. Allowing for
the effects of dust makes little difference to this estimate.

The ionizing radiation pressure from \thOr\ that acts on the
photoevaporated flow can be written as 
\begin{equation}
  \label{prad}
  P_{\rm rad} = \frac{F_\star \langle h \nu
    \rangle}{c}  \ , 
\end{equation}
where $c$ is the speed of light, $F_\star$ is the {\em unattenuated}
ionizing flux from \thOr, and $\langle h \nu\rangle$ is the mean
energy of ionizing photons absorbed in the flow ($\simeq 13.6
\mbox{eV}$). Hence, 
\begin{equation}
  \label{pbalance}
  \frac{P_{\rm rad}}{P_{\rm gas}} \,\simeq\, \frac{h\nu\zero F_\star {\cal
      M}\zero}{\mu m_{\rm H} c\zero c F\zero} \,=\, 0.033\, {\cal
    M}\zero\, 
  \beta_{\%}^{-1} \, = \, 0.022 \, d_{17}^{-1} \, r_{15}^{1/2} \,
  \dot{S}_{49}^{1/2} \ .
\end{equation}
This ratio is always significantly less than unity, therefore the
ionizing radiation pressure is {\bf incapable} of confining the
photoevaporated flow. If there were enough dust opacity at the base of
the flow, then it is conceivable that the non-ionizing radiation from
\thOr\ (FUV, optical) may make a significant contribution to the
radiation pressure. However, the bolometric luminosity of \thOr\ is
only $\sim 3$ times its Ly-c luminosity, so the above conclusion is
unchanged and radiation pressure falls an order of magnitude short of
the thermal pressure even for the closest proplyds ($d_{17} \simeq
0.5$). 

Turning now to the ram pressure, $P_{\rm hyd}$, of the stellar wind from
\thOr, this will be given by
\begin{equation}
  \label{ram}
  P_{\rm hyd} = \rho_{\rm w} v_{\rm w}^2 \ , 
\end{equation}
where $\rho_{\rm w}$ and $v_{\rm w}$ are respectively the stellar wind
density and velocity. Since the wind is radiation-driven, one would
expect the ratio $P_{\rm hyd}/P_{\rm rad}$ to be of order unity and,
using the observed parameters of \thOr\ \cite{How89,Pan73}, one finds
\cite{Hen96} that this is indeed the case. However, although the
radiation pressure must act on the base of the wind where the majority
of the recombinations occur, the ram pressure need not do so, but will
act on the surface of contact between the evaporated flow and the
stellar wind, wherever that may be. Hence, the photoevaporated gas
will flow divergently away from the disk until its pressure (reduced
by geometric dilution) falls to that of $P_{\rm hyd}$, at which point
it can be confined by the stellar wind.

\section{Analytic Two-Wind Models}
\label{sec:anal}
The analytic model \cite{Hen96} depends chiefly on the dimensionless
parameter $\lambda \equiv P_{\rm gas}/P_{\rm hyd}$. From the
discussion of the previous section, it is evident that $\lambda > 1$,
in which case the photoevaporated flow, which is initially mildly
supersonic \cite{Dys68,Kah69,Ber89}, will begin to flow freely away
from the disk. It is assumed that the streamlines are straight and
that the initial flow diverges with a half-opening angle of
45$^\circ$. If the velocity remained constant, the density would fall
as $(1+z)^{-2}$, where $z$ is the height above the disk in units of
the disk radius, but a pressure gradient causes the flow to
accelerate.
  
\begin{figure}[tbp]
  \begin{center}
    \leavevmode
    \psfig{file=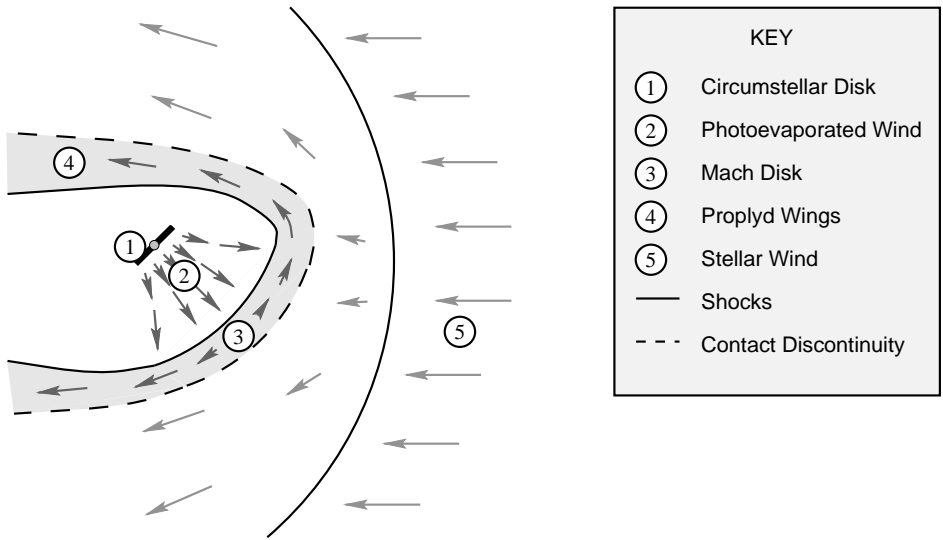,width=\textwidth}
    \caption{Schematic diagram of the two-wind interaction
      model. Ionizing photons from \thOr\ drive a photoevaporated wind
      from one face of the circumstellar disk, which interacts
      supersonically with \thOr's stellar wind.}
    \label{fig:cartoon}
  \end{center}
\end{figure}
The flow will shock at the point where its pressure has fallen 
to that of $P_{\rm hyd}$, which occurs at a distance
\begin{equation}
  \label{bigD}
   D \simeq \frac{1.19\, (\ln\lambda)^{1/4} \lambda^{1/2}}
    {\cos^2\!\theta\zero} \, r_{\rm d} = \mbox{4--20}\, r_{\rm d}\ , 
\end{equation}
where $\theta\zero$ is the inclination angle of the disk normal with
respect to the direction to \thOr\ and, for the second equality,
$\lambda = 10$--100 is assumed. The shock will be radiative, so can be
treated as isothermal, although the radiation from the shock itself
makes a negligible contribution to the proplyd luminosity. A shock
will also form in the wind from \thOr, but this will be non-radiative,
hence the assumption of ram pressure balance used to derive
equation~\ref{bigD} is not strictly valid (see \sect{sec:num}).

A thin, almost flat, layer (Mach disk) of shocked
photoevaporated flow material forms parallel to the circumstellar disk
and gas flows outwards along this layer, reaching a velocity at the edge
of
\begin{equation}
  \label{vel}
  v_\ell \simeq 3.8 (\ln \lambda)^{1/2} \, c\zero = 
  \mbox{70--100\kms} \ . 
\end{equation}
The gas is then swept back by the wind of \thOr\ to form the proplyd
wings and tail. Figure~\ref{fig:cartoon} illustrates the components of
the model in a schematic form.

For reasonable values of $\lambda$, the photoevaporated disk wind is
the brightest component of the model (with a luminosity $\simeq 0.5
\lambda^{1/2}$ times that of the Mach disk plus tail) and also the
smallest, leading to a core-halo morphology \cite[Fig.~11]{Hen96}.

\section{Comparison with Observations}
\label{sec:comp}
The ensemble properties of the proplyds are quite well reproduced by
the analytic model. The models show good agreement with the observed
trends of proplyd size and luminosity vs.\ distance from \thOr, the
former increasing and the latter decreasing
\cite[Figs.~9~and~10]{MCu95,Hen96}. The implied circumstellar disk
radii are between 20 and 60 au ($r_{15} \simeq 0.3$--$1.0$). These
correlations, however, are rather insensitive to the details of the
model.

\begin{figure}[tbp]
  \begin{center}
    \leavevmode
    \psfig{file=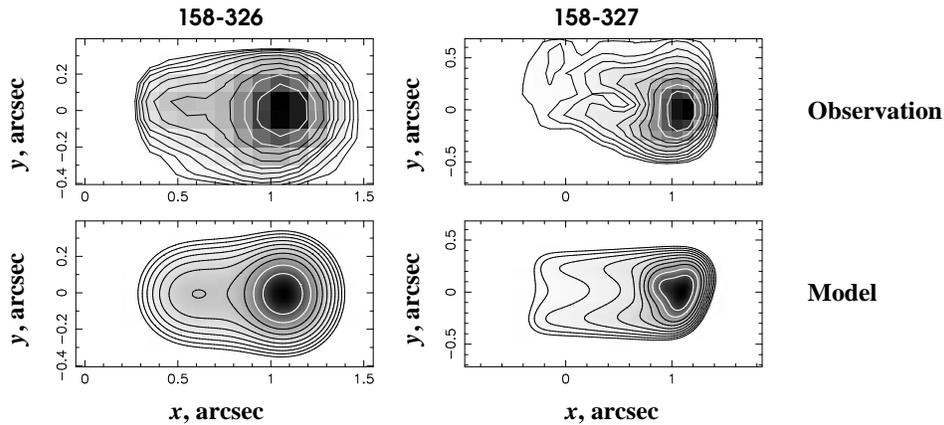,width=\textwidth}
    \caption{Comparison of morphological predictions of the two-wind
      interaction model with observations. Contours and greyscales
      show H$\alpha$ surface brightness (logarithmic scale) for {\em
        HST} observations and model images. The interval between
      successive contours is $2^{1/2}$.}
    \label{fig:comp}
  \end{center}
\end{figure}
The morphologies of individual proplyds are compared with model
predictions in Figure~\ref{fig:comp} (a larger sample is given in
Fig.~14 of \citeauthor{Hen96}, \citeyear{Hen96}), where it can be seen
that the models successfully reproduce single and double tails, both
of which are observed \cite{ODe94,ODe96,Joh96}.  However, the double
tails may be merely the result of absorption in the core of the tail,
which is not consistent with the models as they stand. The crescent
head observed in many proplyds would correspond to the Mach disk in
the models, but this is rather problematic since the models predict
that this should be less bright than the photoevaporated wind
component (\sect{sec:anal}), which is not the case for most proplyds,
although dust absorption at the base of the wind ($\tau\simeq\,$0.5--1
for the closest proplyds) would alleviate this problem (this is
included in the fit to OW~158--327).

Detailed comparisons between model predictions and high resolution
\Othree\ spectra of individual proplyds are presented in
\citeauthor{Hen97} \shortcite{Hen97}. The evaporated wind produces the
bright core of the line, with width of a few times the sound speed in
the ionized gas, while the Mach disk and tail produce the
high-velocity line wings that are observed. However, in order to
reproduce the $\simeq 100 \kms$ widths of the line wings seen in LV~5
(\P5) and LV~2 (\P2), values of $\lambda = 50$--200 are required,
which are 3--4 times larger than those found in fitting the
morphologies of the same objects \cite{Hen96}.

\section{Hydrodynamical Simulations}
\label{sec:num}
\begin{figure}[tbp]
  \psfig{file=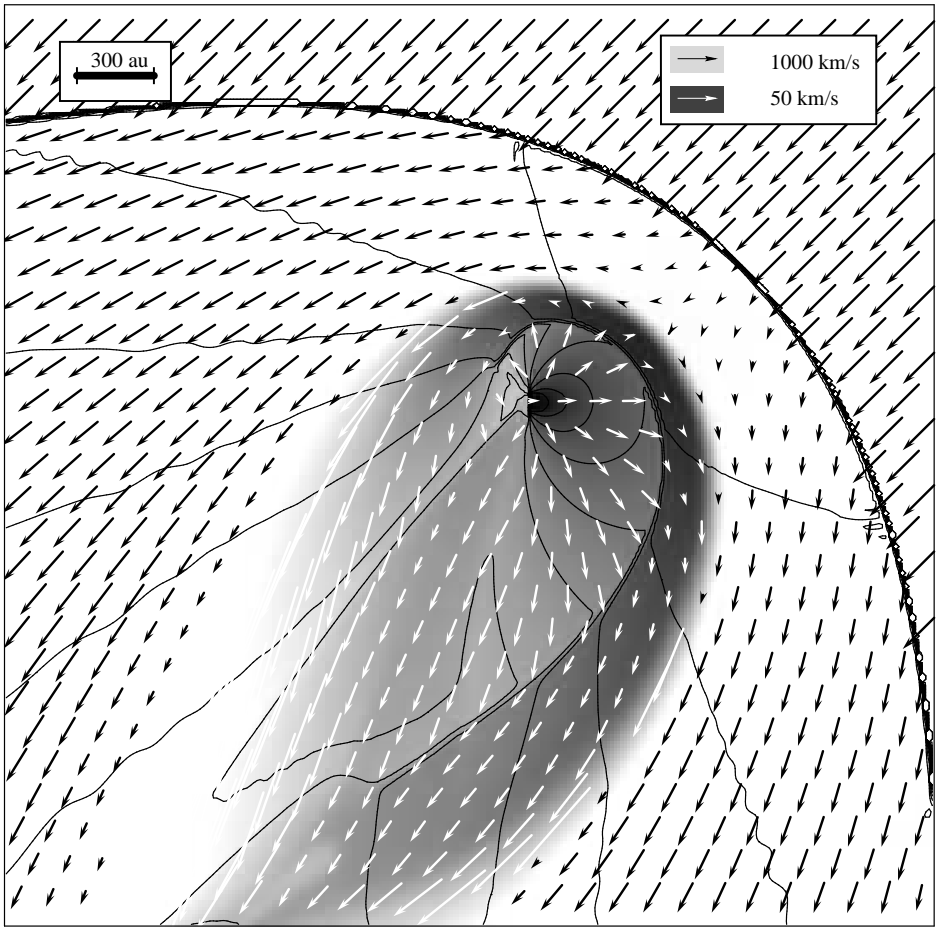,width=\textwidth}
  \caption{Hydrodynamical simulation of the two-wind
    interaction, calculated in 2D slab symmetry with a grid size of
    $300\times 300$ cells.  Greyscale shows the gas density while
    contours show the gas pressure (both logarithmic scale). Arrows
    show gas velocity. Photoevaporated disk material (white arrows)
    has an isothermal equation of state. Stellar wind material
    (black arrows) has an adiabatic equation of state. Note the
    different velocity normalizations of the black and white arrow
    lengths. The parameters of the model are $r_{\rm d} = 30$~au,
    $\theta\zero = 45^\circ$, $n\zero = 2 \times 10^5
    \mbox{cm}^{-3}$, $n_{\rm w} = 1 \mbox{cm}^{-3}$, $u_{\rm w} =
    1000\kms$.}
  \label{fig:hydro}
\end{figure}
Figure~\ref{fig:hydro} shows the results of an example numerical
simulation of the two-wind interaction \cite{Hen97b}. In this
simulation, the circumstellar disk (oriented vertically in the figure)
is inclined by $45^\circ$ with respect to the direction of the stellar
wind from \thOr\ (other parameters are described in the figure
caption). The transfer of ionizing radiation in the photoevaporated
flow is not calculated self-consistently in the models, but the
boundary conditions are assumed constant over the disk surface and are
taken from the analytic model. Also, the simulation parameters
correspond to a rather small value of $\lambda$ ($\simeq 8$), since a
larger value would require an unfeasibly large computational grid.

The main differences with respect to the analytic calculation are due
to the relaxation of two arbitrary assumptions of the model. Firstly,
the photoevaporated flow is calculated self-consistently, instead of
being assumed to follow straight streamlines with opening angle
$45^\circ$. Pressure gradients in the mildly supersonic wind in fact
cause the flow to diverge increasingly with distance from the disk.
Secondly, the (non-radiative) shock in the stellar wind is treated
properly, instead of merely assuming ram pressure balance between the
two winds as was done in the analytic model. Both these factors affect
the shape of the dense layer of shocked photoevaporated wind material,
which is more curved than in the analytic calculation. The flat ``Mach
disk'' of the analytic model is no longer apparent and there is no
sharp distinction between the Mach disk and tail. The gas velocities
reached in the shocked photoevaporated wind are also slightly smaller
than in the analytic model. Unfortunately, the use of slab symmetry,
which allows the asymmetric interaction to be modelled in two
dimensions, means that it is not possible to produce emission maps or
spectra from the simulations. Nonetheless, the morphology and
kinematics of the simulations are rather similar to those of the
analytic calculation modulo the differences noted above.  The
arguments for and against the two-wind interaction model are hence
little affected.

\section{Discussion and Speculation}
\label{sec:discuss}
Despite the success of the two-wind interaction models in reproducing
the observed morphologies and kinematics of the proplyds, some
problems remain. In particular, the apparent absorption in the tails
of some objects and the [\ion{O}{3}]/IR arcs \cite{p:Bal95,Hay94} that
are seen between the closer proplyds and \thOr\ are both hard to
explain with the two-wind model. An alternative view
\cite{p:Bal95,Joh96} is that the disk evaporation is controlled by
non-ionizing FUV photons, with the ionization front occurring away
from the disk. The proplyd morphology would then be determined by the
shape of the ionization front. The hydrodynamic interaction with the
stellar wind from \thOr\ would, on this view, still occur, but farther
out in the flow, perhaps producing the [\ion{O}{3}]/IR arcs.  This
model has had most success in explaining the characteristics of HST~10
(OW~182--413), but this object does not seem to be typical of proplyds
as a class (in particular, its tail does not point exactly away from
\thOr\ and it may not contain a central star).  \citeauthor{Joh96}
\shortcite{Joh96} compare the mass-loss rates from photodissociated
and photoionized disk winds and conclude that the former will dominate
for all proplyds. However, they {\em assume} that the warm ($\simeq
1000\,$K) photodissociated gas will be able to flow freely away from
the disk at its sound speed ($\simeq 2.5\kms$), but this is not
necessarily the case.

If one allows, for the sake of argument, that a free-flowing
photodissociated wind, with a density at its base of $10^6
n_{{\rm n},6}\,\mbox{cm}^{-3}$, is initially established, then, once the
ionizing radiation from \thOr\ is switched on, an R-type ionization
front will be driven rapidly into the flow. For proplyds closer than
\begin{equation}
  \label{zap}
  d^\prime_{17} \simeq 0.55\, n_{{\rm n},6}^{-1}\, r_{15}^{-1/2}\,
  \dot{S}_{49}^{1/2} \ ,  
\end{equation}
the flow will be immediately ionized all the way down to the disk. For
proplyds further away, the ionization front undergoes a transition to
D-type some way out from the disk, at which point its progress will
slow and it will begin to drive a shock into the atomic flow. The
density of newly ionized gas $n_0$ will adjust itself to that given by
equations~\ref{base} and~\ref{beta} of \sect{sec:constraints}, but
with $r_{\rm d}$ replaced by the radius of the ionization front. 
The pressure of the ionized gas will be roughly 20 times that of
neutral gas of the same density, so that for proplyds closer than
\begin{equation}
  \label{stall}
  d^{\prime\prime}_{17} \simeq 20\, n_{{\rm n},6}^{-1}\, r_{15}^{-1/2}\,
  \dot{S}_{49}^{1/2} \ ,
\end{equation}
the shock will reach the surface of the disk before stalling, hence
quenching the neutral flow in a time of 30--500~years. In proplyds
farther away than $d^{\prime\prime}_{17}$ from \thOr, the shock will
stall at a distance $z_0 r_{\rm d}$ from the disk, where
\begin{equation}
  \label{zstall}
  1 + z_0 \simeq 0.136\, n_{{\rm n},6}^{2/3} \, d_{17}^{2/3}\,
  r_{15}^{1/3}\, \dot{S}_{49}^{-1/3}  \ .
\end{equation}
The chief uncertainty in the estimates of $d^\prime_{17}$ and
$d^{\prime\prime}_{17}$ is the density at the base of the neutral
flow. However, taking the parameters of HST~10 \cite{Joh96}, $r_{15} =
1.3$, $z_0 = 2.3$, $d_{17} = 5$, and assuming $n_{{\rm n},6}$ is the
same for all proplyds, one finds that \mbox{$d^\prime_{17} \simeq
  0.03\, r_{15}^{\scriptscriptstyle -1/2}$} and
\mbox{$d^{\prime\prime}_{17} \simeq r_{15}^{\scriptscriptstyle
    -1/2}$}. No proplyds are observed with \mbox{$d_{17} <
  d^\prime_{17}$} but a substantial fraction have \mbox{$d_{17} <
  d^{\prime\prime}_{17}$}, although the exact number depends on the
distribution of disk radii. This can only be determined directly for
the dark silhouette disks \cite{McC96}, which show $r_{15} =
0.4$--7.6, but the bright proplyds are likely to have smaller disks
($r_{15} \simeq 0.1$--1.0, \citeauthor{Hen96}, \citeyear{Hen96};
\citeauthor{Joh96}, \citeyear{Joh96}). Hence, roughly half of all
bright proplyds will {\bf not} have an extended neutral evaporated
flow.

The real situation is undoubtedly much more complicated than portrayed
above (c.f.\ \citeauthor{Ber96}, \citeyear{Ber96}), but the basic
argument, that the neutral flow must have a higher pressure than the
overlying ionized flow in order to exist, should remain valid. A
further problem for the neutral flow is gravity: the escape speed from
the circumstellar disk will equal the sound speed in the neutral gas
at a disk radius of $r_{{\rm esc}, 15} \simeq 2.1 M_\star$, where
$M_\star$ is the mass of the central star in solar masses
($\simeq\,$0.1--2, \citeauthor{McC94}, \citeyear{McC94}). Hence,
except for the proplyds with the lowest mass central stars, gravity
will dominate the dynamics of the photodissociated region. 

Note that the argument against radiation pressure in
\sect{sec:constraints} applies a fortiori to a neutral
photodissociated flow since its pressure would have to be larger than
the ionized flow.  However, the confinement problem could be
circumvented if it were maintained that the material in the tail,
instead of having been redirected from an initial flow towards \thOr,
was instead part of a flow from the back side of the disk, possibly
driven by the diffuse radiation field. This could also explain the
absorption seen in the core of some tails, but whether the flow would
be dense enough for this is not clear. Alternatively, the tails could
be formed from the remnants of a dense slow wind from a massive star
\cite{Sut97}.

In conclusion, the two-wind interaction model has had qualified
success in explaining the observed properties of the proplyds closer
to \thOr. Various discrepancies remain, however, and further work is
necessary both in extending this model and in developing
alternatives. 

\vspace{0.2cm}

{\bf Acknowledgements:} We are very grateful to Alex Raga, Susana
Lizano and John Meaburn for their contributions to the work discussed
in this paper. WJH also acknowledges useful discussions with John
Dyson, Dave Hollenbach, Mark McCaughrean and Bob O'Dell. Financial
support for this research has been provided by DGAPA-UNAM under
project number IN105295 and by Cray Research, Inc.

\end{document}